\documentclass[aps,pra,twocolumn,a4paper,showpacs,superscriptaddress,floatfix,10pt]{revtex4}

\usepackage{graphicx}
\usepackage{amsmath}
\usepackage{epsfig}
\usepackage{helvet}
\usepackage{amssymb}
\usepackage{epstopdf}
\usepackage{color}
\usepackage{xspace}
\usepackage{bbold}

\newcommand{\bra}[1]{\mbox{$\langle #1 |$}}
\newcommand{\ket}[1]{\mbox{$| #1 \rangle$}}

\newcommand{\tr}{\mbox{tr}}
 \newcommand{\LIOM}{IOM\xspace}

\def\ZZ{\mathcal{Z}}
\def\PP{\mathcal{P}}

\def\lz{\ell_{\mathcal{Z}}}

\begin{document}

\title{Spectral tensor networks for many-body localization}

\author{A. Chandran}
\author{J. Carrasquilla}
\author{I. H. Kim}
\author{D. A. Abanin}
\author{G. Vidal}
\affiliation{Perimeter Institute for Theoretical Physics, Waterloo, Ontario N2L 2Y5, Canada}  

\date{\today}

\begin{abstract}
Subsystems of strongly disordered, interacting quantum systems can fail to thermalize because of the phenomenon of many-body localization (MBL).
In this article, we explore a tensor network description of the eigenspectra of such systems.
Specifically, we will argue that the presence of a complete set of local integrals of motion in MBL implies an efficient representation of the entire spectrum of energy eigenstates with a single tensor network, a \emph{spectral} tensor network.
Our results are rigorous for a class of idealized systems related to MBL with integrals of motion of finite support.
In one spatial dimension, the spectral tensor network allows for the efficient computation of expectation values of a large class of operators (including local operators and string operators) in individual energy eigenstates and in ensembles.
 \end{abstract}


\maketitle


\emph{Introduction: }
It is widely believed that isolated interacting quantum systems evolving under their own Hamiltonian act as their own bath, so that finite subsystems are described by equilibrium statistical mechanics at late times irrespective of the global state \cite{Deutsch:1991ss,Srednicki:1994dw,Polkovnikov:2011ys}.
A notable exception is the Anderson insulator \cite{Anderson:1958ly} and its interacting generalization: the many-body localized insulator \cite{Anderson:1958ly,Basko:2006aa, Gornyi:2005lq,Altman:2014la,Nandkishore:2014nx}.
Such insulators occur in systems with strong quenched disorder.
The absence of transport in these systems prevents equalization of spatial imbalances of conserved densities and blocks thermalization \cite{Basko:2006aa, Gornyi:2005lq,Pal:2010gs,Monthus:2010vn,Berkelbach:2010aa,Luca:2013qe}.
Phenomenologically, the failure to thermalize is captured by an extensive number of \emph{spatially local} integrals of motion (\LIOM) that retain local memory of the initial state for infinite time \cite{Serbyn:2013rt,Huse:2013kq,Ros:2014nr,Imbrie:2014jk}.
The presence of the local \LIOM also predicts that 1) all the eigenstates of the Hamiltonian have very low entanglement (area law) \cite{Bauer:2013rz,Kjall:2014ly,Friesdorf:2014ib} 2) the entanglement entropy grows only logarithmically with time under a global quench \cite{Bardarson:2012kl,Znidaric:2008aa,Vosk:2013sf,Huse:2013kq,Serbyn:2013xe,Nanduri:2014pi}, 3) systems with MBL can support order forbidden by statistical mechanics at finite energy density \cite{Huse:2013rc,Bahri:2013qr,Chandran:2014eu}.
The local \LIOM are thus believed to capture all the properties of many-body localization (MBL) in the absence of a mobility edge.

The area law of entanglement entropy obeyed by the highly excited eigenstates and the existence of local \LIOM suggests the tensor network formalism as a natural framework for studying MBL \cite{Hastings:2007kx,Verstraete:2008ve}.
The goal of this article is to make this connection precise. 
Specifically, we will argue that the presence of a complete set of local \LIOM implies an efficient representation of the entire spectrum of energy eigenstates with a single tensor network, a \emph{spectral tensor network}. 
This is a surprising result, which is not known to hold for any other quantum phase of matter \footnote{ Such spectral tensor networks have previously been constructed only for very special systems such as Kitaev's toric code \cite{Kitaev:2003vn} and free fermions \cite{Ferris:2014qa} which require fine tuned Hamiltonians.}.
In one spatial dimension, the spectral tensor network allows for the efficient computation of expectation values of a large class of operators (including local operators and string operators) in individual energy eigenstates and in ensembles. 
When the set of IOM is incomplete, as is the case with topological order or possibly with a mobility edge, the resulting tensor network efficiently represents projectors onto subspaces with well-defined eigenvalues for each of the \LIOM.

The \LIOM in the physical setting of MBL have exponentially suppressed tails away from a central finite region, that is, they are \emph{quasi-local} integrals of motion (q\LIOM).
In the first half of this article, we neglect the tails and work with \emph{strictly local} integrals (s\LIOM) with a strictly finite region of support. 
We show that (1) each s\LIOM can be efficiently represented as a tensor network and (2) the number of s\LIOM with overlap on any given site is finite, even in the thermodynamic limit.
We then rigorously prove that (1) and (2) imply the existence of an efficient spectral tensor network. 
In the second half, we consider the generic case of q\LIOM. 
We argue that the q\LIOM satisfy analogs of (1) and (2) above and present an approximate spectral tensor network representation. 
We then provide numerical evidence for (1) and (2) in the q\LIOM case, using free fermions for (2) \footnote{Free fermions are not special in the tensor network formalism as the formalism is agnostic to interactions.}.

Our results, which assume the ability to efficiently identify the q\LIOM, open the path to the efficient (and thus scalable) numerical simulation of MBL systems.
With the much larger system sizes accessible (especially in one dimension), questions about the nature of MBL at intermediate disorder and the scaling theory near the localization-delocalization transition can be addressed.
We therefore expect the tensor network formalism to become an important tool in the study of MBL. 

\emph{Setting: }
Consider a lattice of $k$-state spins with $N$ sites in $D$ spatial dimensions.
For simplicity, we assume that the lattice is a hypercube and that $k=2$, although our results generalize readily to any lattice and to any integer $k$.
Let the Hamiltonian of the system be denoted by $H$ and the set of IOM be denoted by $\{\ZZ^{(i)}\}, i=1,\ldots N_{\ZZ}$.
By definition, the IOM commute with $H$. Our first assumption is that they also commute with one another:
\begin{align}
\label{eq:integrals}
[H, \ZZ^{(i)}]=0, \qquad [\ZZ^{(i)},\ZZ^{(j)}]=0.
\end{align}

Second, we assume that the set $\{\ZZ^{(i)}\}$ is algebraically independent. 
That is, we assume that any linear combination of the tensor products of the IOM in the set is an IOM that does not belong to the set $\{\ZZ^{(i)}\}$.
Roughly speaking, this requirement prevents redundancy in the given set $\{\ZZ^{(i)}\}$ of IOMs.

Third, we assume that the IOMs are local. 
The quasi-local IOM are localized about a central core with exponentially suppressed tails, while the strictly local ones have no tails.
A quantitative measure of quasi/strict locality is through the Frobenius norm of the partially traced IOM.
The Frobenius norm (hitherto norm) of an operator $O$ is defined as $\| O\|= \sqrt{\textrm{Tr} (O^\dagger O)/2^{N-1}}$.
Let $S$ be the hypercube of linear size $\lz$ that denotes the central (full) support of the q\LIOM (s\LIOM).
Let $A$ be a hypercube of length $\ell_A$ that encloses $S$, $S\subset A$.
The spatial locality of the \LIOM implies that the full IOM and the partially traced \LIOM over $\bar{A}$, the complement of region $A$, are close as measured by the norm:
\begin{align}
\label{Eq:sqLIOMdef}
\left|\left| \ZZ^{(i)}  - \textrm{Tr}_{\bar{A}} \ZZ^{(i)}\otimes \frac{\mathbb{1}_{\bar{A}} }{2^{N_{\bar{A}}} } \right|\right|  \begin{cases}
\leq \exp(-\ell_A/\xi) &\ZZ^{(i)} \mbox{ is q\LIOM} \\
= 0 & \ZZ^{(i)} \mbox{ is s\LIOM}.
 \end{cases}
 \end{align}
Above, $N_{\bar{A}}$ is the number of spins in $\bar{A}$ and $\xi$ is the length scale over which the exponential tail of the q\LIOM decays (a localization length).

Finally, we assume that the \LIOM are orthogonal to one another.
That is, $\textrm{Tr} \ZZ^{(i)} \ZZ^{(j)} = 0 $ if $i\neq j$.
We will require this property when we argue for an approximate spectral tensor network in the quasi-local case.

\emph{Examples of systems with local \LIOM: }
The simplest example of a system with s\LIOM satisfying the above conditions is the diagonal Hamiltonian:
\begin{align}
\label{Ham:Exdiagonal}
H = \sum_i h_i \sigma^z_i + J_{i} \sigma^z_i \sigma^z_{i+1}
\end{align}
where $h_i, J_i$ are identical and independent random variables drawn from the box distribution of width $W$ in dimension $D=1$.
$H$ is trivially MBL \cite{Huse:2013kq, Swingle:2013pb}. 
This can be seen in a number of ways.
For example, the eigenstates of $H$ are product states in the $z$ basis and do not reproduce expectation values in the Gibbs ensemble at any temperature.
Also, as all the terms in $H$ commute, the Kubo formula for thermal conductivity is zero at any temperature and there is an absence of energy transport.
One set of s\LIOM in this system is $\ZZ^{(i)}= \sigma^z_i$.
Each $\sigma^z_i$ has support only on site $i$, the $\sigma^z_i$ operators on different sites commute with one another and with $H$, and the set is clearly algebraically independent and orthogonal.
Thus, all the conditions we required above are satisfied.

A system with q\LIOM can be constructed from Eq.~\eqref{Ham:Exdiagonal} by conjugating $H$ with a quasi-local unitary $U$ \cite{Swingle:2013pb}.
An example of a quasi-local $U$ is the time evolution operator of some local Hermitian operator $W$ for fixed time.
The new Hamiltonian: $H' = U H U^\dagger$ has the same eigenvalues as $H$; the eigenstates are obtained by the action of $U$ on the $z$-product states.
The IOMs are now: $\ZZ^{(i)} = U \sigma^z_i U^\dagger$; they inherit the mutual commutativity, algebraic independence and orthogonality from the s\LIOM set $\{\sigma^z_i\}$.
The quasi-locality of $U$ further guarantees that the $\ZZ^{(i)}$ are quasi-local by Lieb-Robinson type arguments \cite{Lieb:1972aa}.

\emph{The spectral tensor network for s\LIOM: }
Consider a system with s\LIOM.
We show below that (1) each s\LIOM can be efficiently represented as a tensor network and (2) the number of s\LIOM with overlap on each site is finite in the thermodynamic limit.
We then rigorously prove that (1) and (2) imply the existence of an efficient spectral tensor network.

\begin{figure}[tbp]
\begin{center}
\includegraphics[width=7cm]{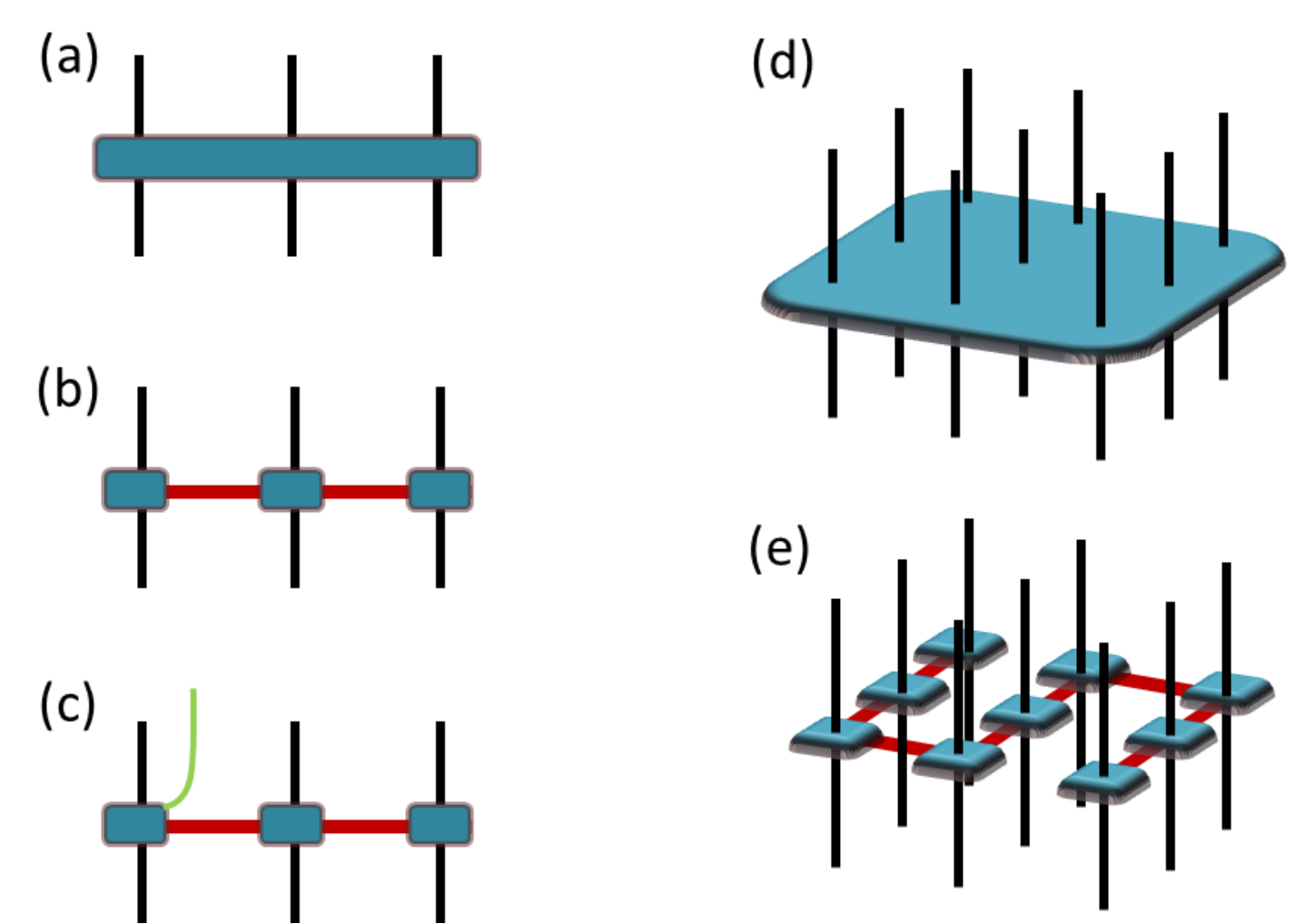}
\caption{ (a) A projector $\PP^{(n)}_{\mu_n}$ in $D=1$ acting on $\ell_{\ZZ}=3$ sites. (b) $\PP^{(n)}_{\mu_n}$ expressed as a MPO. (c) All $q$ projectors $\{\PP^{(n)}_{\mu_n}\}$ represented as an MPO with additional index $\mu_n$. (d) A projector $P^{(n)}_{\mu_n}$ in $D=2$ acts non-trivially on $(\ell_{\ZZ})^{2} = 9$ sites. (e) The projector in (d) expressed as a MPO. The bond dimension is all cases is independent of system size.
}
\label{fig:SpectralTN2}
\end{center}
\end{figure}

\emph{(1) Efficient representation of each s\LIOM:}
The eigenvalue decomposition of each s\LIOM is:
\begin{equation}\label{eq:eigenP}
    \ZZ^{(n)} = \sum_{\mu_n=1}^q \lambda^{(n)}_{\mu_n} \PP^{(n)}_{\mu_n},
\end{equation}
where $\lambda^{(n)}_{\mu_n}$ are the $q$ distinct eigenvalues of $\ZZ^{(n)}$ and the operators $\PP^{(n)}_{\mu_n}$ are orthogonal projectors that add to unity.
Using standard matrix product techniques, each projector $\PP^{(n)}_{\mu_n}$ can be expressed in a MPO form \cite{Verstraete:2008ve,Schollwock:2011qf} (Fig.~\ref{fig:SpectralTN2}). 
This requires choosing a path or \textit{snake} in the support of $\ZZ^{(n)}$ and, for each bipartition, performing a singular value decomposition. 
A MPO for the whole set of $q$ projectors (Fig. \ref{fig:SpectralTN2}(c)) is then obtained by attaching an extra index with $q$ values through the whole snake.
As the integrals are s-local, the size of the representation is independent of system size and is efficient.

\begin{figure}[tbp]
\begin{center}
\includegraphics[width=8.5cm]{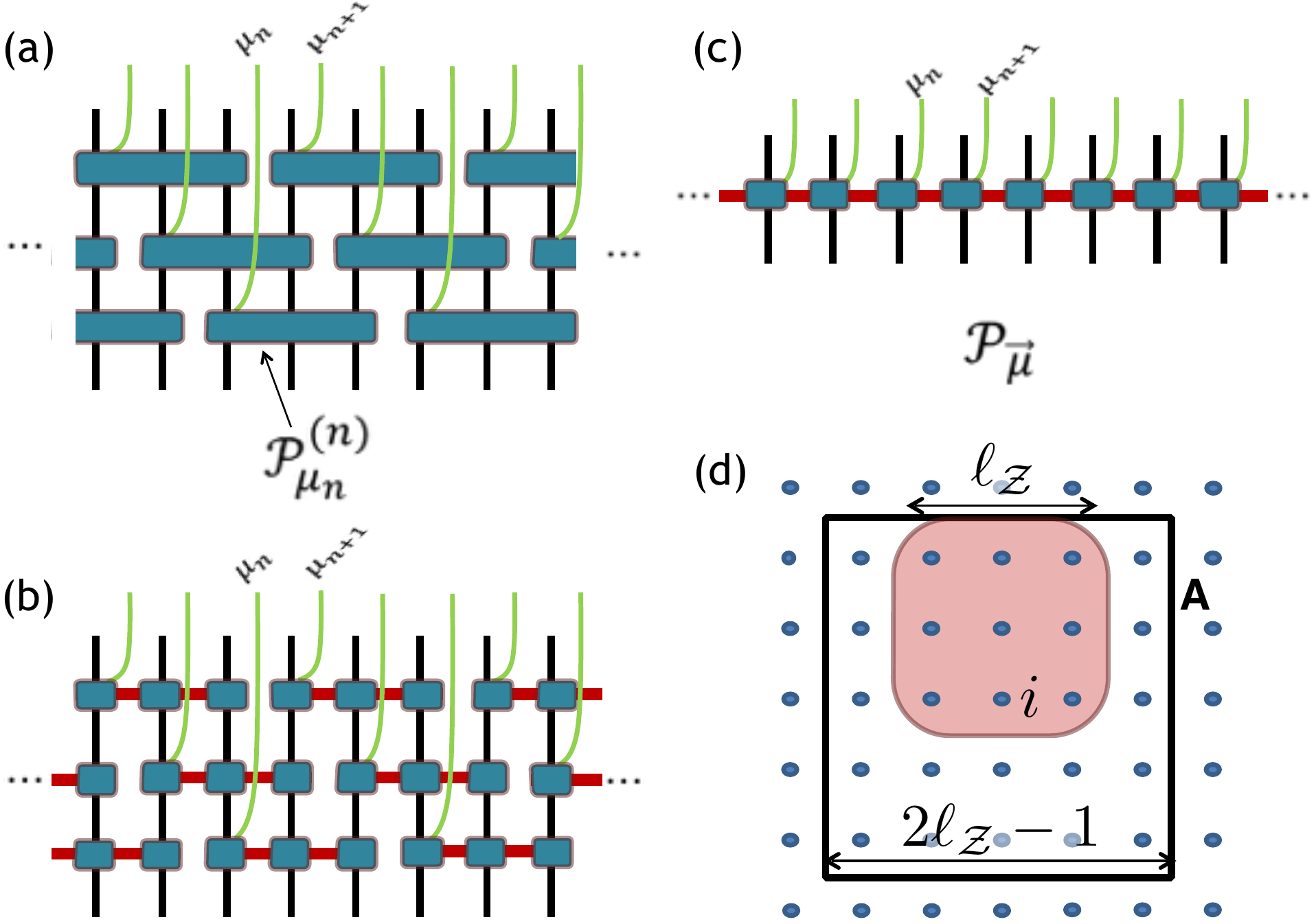}
\caption{ (a)-(c) Construction of the spectral tensor network in $D=1$ ($N_{\ZZ}=N$). (a) Product of $N$ projectors $\{\PP^{(1)}_{\mu_1},\cdots, \PP^{(N)}_{\mu_N}\}$. (b) Each projector decomposed as a MPO as in Fig.~\ref{fig:SpectralTN2}. (c) Product of MPOs in (b) produces the spectral tensor network for $\PP_{\vec{\mu}}$. The bond dimension in all cases is independent of $N$. (d) The number of s\LIOM overlapping with site $i$ is upper bounded by the number of sites in $A$.}
\label{fig:SpectralTN3}
\end{center}
\end{figure}

\emph{(2) Finite number of s\LIOM at each site: }
For ease of visualization, let $D=2$, as in Fig.~\ref{fig:SpectralTN3} (d).
Let $N(i)$ denote the number of s\LIOM that overlap with site $i$. 
We design the square $A$ of length $(2\ell_{\ZZ}-1)$ with site $i$ at the centre so that all the s\LIOM that overlap with $i$ are entirely contained in $A$ (e.g. s\LIOM with support in pink square in Fig.~\ref{fig:SpectralTN3} (d)). 
Thus, $N(i)$ is upper bounded by the number of s\LIOM entirely contained in $A$. 
As the set of s\LIOM is algebraically independent, the number of s\LIOM entirely in $A$ is at most equal to the number of sites in $A$.
Therefore:
\begin{align}
\label{Eq:Ni}
N(i) \leq (2 \ell_{\ZZ} -1)^D
\end{align}
The upper bound on $N(i)$ is independent of system size $N$, hence the number of s\LIOM at each site is finite.
Note that the bound is not tight: in general, $q \geq 2 $ (Eq.~\eqref{eq:eigenP}) and $N(i)$ is at most $(2 \ell_{\ZZ} -1)^D/\log_2 q$.

\emph{Efficient spectral tensor network: }
We now prove that (1) and (2) above imply an efficient spectral tensor network.
Notice that Eq.~\eqref{eq:integrals} implies that eigenvectors of $H$ can be labelled by their collection of eigenvalues under $\ZZ^{(n)}$.
If the number of s\LIOM in the set is equal to $N$, then the labelling identifies a single eigenstate, else it only identifies a subspace.
In either case, the projector onto an energy subspace labelled by the eigenvalues $\mu_1, \ldots \mu_{N_{\ZZ}}$ of the s\LIOM $\ZZ_1\ldots \ZZ_{N_{\ZZ}}$ is equal to the product of the appropriate projectors of the s\LIOM:
\begin{align}
\PP_{\vec{\mu}}  = \PP_{\mu_1}^{(1)} \PP_{\mu_2}^{(2)} \cdots \PP_{\mu_{N_{\ZZ}}}^{(N_{\ZZ})}.
\end{align}
The tensor network representation for $\PP_{\vec{\mu}}$ follows by multiplying together the MPOs for the $N_{\ZZ}$ projectors.
If we now allow the $\mu_n$ indices to take all possible values, we obtain the promised spectral tensor network.

The spectral tensor network and the steps in its construction are shown in Fig.~\ref{fig:SpectralTN3}(a)-(c).
The network has $2N+N_{\ZZ}$ open indices: $N$ incoming and $N$ outgoing physical indices shown in black, and $N_{\ZZ}$ additional open indices $\mu_n$ shown in green that identify the projector $\PP_{\vec{\mu}}$ \footnote{For simplicity, we have assumed that $N_{\ZZ}=N$ and attached one $\mu_n$ index to each tensor or lattice site. There is no unique way to assign $\mu_n$ indices to sites. However, as the number of \LIOM with overlap on a site is independent of $N$, at most a constant number of $\mu_n$ will be assigned to a single site as $N\rightarrow \infty$.}.
The virtual bonds are shown in red.

Finally, we prove that the spectral tensor network is efficient, that is, that the bond dimension is finite in the thermodynamic limit.
We have already shown that the representation of each $\PP^{(n)}_{\mu_n}$ is efficient.
From Fig.~\ref{fig:SpectralTN3}, it is easily seen that we further need the number of s\LIOM that overlap with a given site to be finite as $N\rightarrow \infty$.
This is result (2) above. 
Thus, the representation is efficient.

\emph{The spectral tensor network for q\LIOM: }
Guided by the intuition that the q\LIOM are roughly like s\LIOM with tails, we argue that approximate versions of (1) and (2) above hold in a system with q\LIOM, implying the existence of an approximate efficient spectral tensor network.

First, we conjecture that the projectors onto q\LIOM can be efficiently represented as tensor networks ((1) above).
This is reasonable based on the definition in Eq.~\eqref{Eq:sqLIOMdef}.
Eq.~\eqref{Eq:sqLIOMdef} implies that the error is controlled by $\ell/\xi$ if we truncate the q\LIOM at some distance $\ell$ greater than $\xi$ or $\ell_{\ZZ}$ from its center.
The truncated object has finite support and hence an efficient MPO representation.
Increasing $\ell$ should only weakly perturb the bond dimension near the centre; thus we expect an efficient representation of the exact q\LIOM even in the limit of $\ell$ approaching the system size.
Note that to ensure that we have an efficient representation of the projector onto \emph{any} eigenstate of $\ZZ^{(i)}$, we need the number of eigenvalues of $\ZZ^{(i)}$ to be finite in the thermodynamic limit. 
In the next section as well as in the physical situation of MBL, this holds.

Next, we need to argue that the number of q\LIOM with overlap on each site is finite as $N\rightarrow \infty$ ((2) above).
This cannot be an exact statement, as each q\LIOM has some weight on every site.
However, we expect that with controllably small error, we can approximate each q\LIOM with an operator of finite support by truncating in real space (see example below).
The truncated q\LIOM are also approximately orthogonal to one another so that $N(i)$ defined in (2) is again upper bounded by the number of sites in $A$ (Fig.~\ref{fig:SpectralTN3} (d)).

The analogs of (1) and (2) above imply the existence of an efficient, approximate spectral tensor network in a system with q\LIOM.

\emph{Numerical evidence for (1) and (2) in q\LIOM system: }
\begin{figure}[tbp]
\begin{center}
\includegraphics[width=7.2cm]{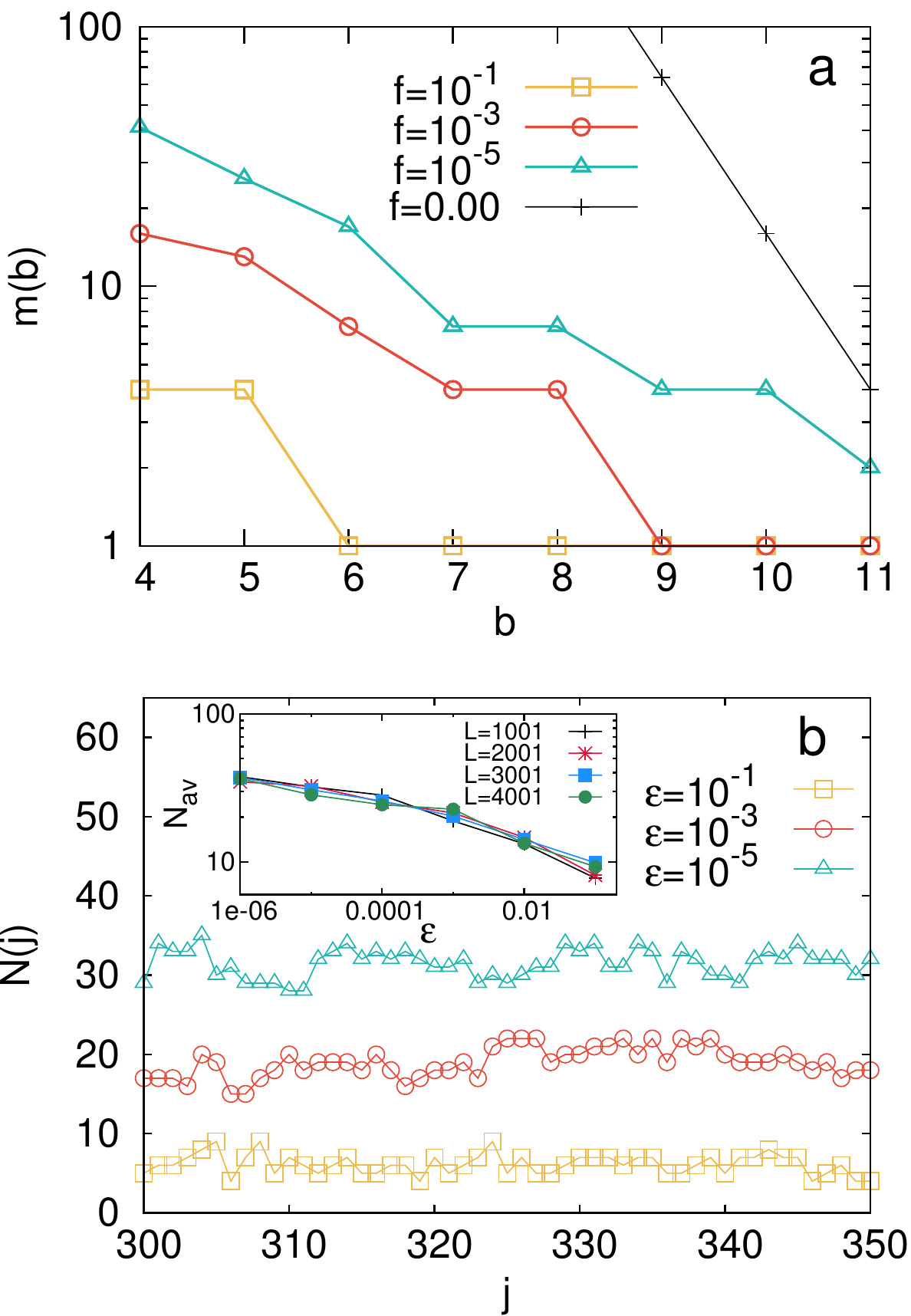}
\caption{ (a) Bond dimension of the MPO representation of $e^{iH_{XYZ}} \sigma_{4}^z e^{-iH_{XYZ}}$ versus the bond index for truncation parameter $f=10^{-1},10^{-3}, 10^{-5}$, $L=12$. 
(b) Number of truncated q\LIOM overlapping with site $j$, $N(j)$, vs site index $j$ in a typical disorder realization of free fermions with $W/t=3, L=1001$ for truncation $\epsilon = 10^{-1}, 10^{-3}, 10^{-5}$. Inset: Averaged value of $N(j)$ vs $\epsilon$ for $L=1001,2001,3001,4001$.}
\label{Fig:Numerics}
\end{center}
\end{figure}
To check if (1) holds, consider a spin-1/2 system of length $L$ with open boundary conditions. 
We create a quasi-local operator $\ZZ$ by conjugating $\sigma_{4}^z$ in the bulk of the chain by a quasi-local unitary $U=e^{-iW}$, where $W$ is a local translationally invariant Hermitian operator.
We then represent $\ZZ$ by a MPO and truncate the representation by discarding all the Schmidt values that are smaller that $f \lambda_{max}$, where $\lambda_{max}$ is the maximum Schmidt value on each bond and $f$ is a small number.
In Fig.~\ref{Fig:Numerics}(a), we show the bond dimension of the truncated MPO representation $m(b)$ versus the bond index $b$ for different $f$ at $L=12$.
Note that the bond between sites $i$ and $i+1$ is indexed by $i$.
First, we find that $m(b)$ quickly approaches one away from site $4$ showing that the truncated $\ZZ$ has finite support.
Second, we find that $m(b)$ is much smaller than the maximum possible value shown in black ($f=0$), suggesting that the MPO representation is efficient and that $m(b)$ is independent of $L$.
We explicitly check that $m(b)$ is independent of $L$ at $f=10^{-3}$ in Appendix~\ref{App:mbvsL}.
Figs.~\ref{Fig:Numerics}(a) and \ref{Fig:Ldep} thus provide evidence for an efficient MPO representation of a quasi-local operator.

To check (2), we consider a system of free fermions in one dimension with nearest neighbour hopping $t$, and random on-site potential uniformly drawn from $[-W, W]$.
The system is fully localized, and the quasi-local \LIOM $\ZZ^{(i)} = \hat{n}^{(i)}$, where $\hat{n}^{(i)}$ is the occupation number operator of the $i$th single-particle eigenfunction. 
Let the $i$th single-particle eigenfunction be localized about site $R_i$.
We truncate each $\ZZ^{(i)}$ to a finite range $\tilde{\ZZ}^{(i)}$ about $R_i$ such that $||\tilde{\ZZ}^{(i)}|| = 1-\epsilon$ (note that $||\ZZ^{(i)}||=1$).
We then count the number of truncated q\LIOM that overlap with site $j$, $N(j)$. 
In the main plot in Fig.~\ref{Fig:Numerics}(b), we plot $N(j)$ vs site index $j$ in a typical disorder realization for $\epsilon=10^{-1}, 10^{-3}, 10^{-5}$ with $W/t=3, L=1001$.
We notice that $N(j)$ is much smaller than $L$, even at small error in norm $\epsilon = 10^{-5}$.
In the inset, we plot the averaged value of $N(j)$ over the entire system, $N_{\textrm{av}}$, vs $\epsilon$ for different $L$. We see that $N_{\textrm{av}}$ is independent of $L$ at all $\epsilon$.
Thus, up to any desired precision, the number of q\LIOM with overlap at each site is finite in the limit $L\rightarrow \infty$.

\emph{Discussion: }
We have presented two results in this article.
First, we rigorously proved the existence of an efficient spectral tensor network in systems endowed with an algebraically independent set of strictly local integrals of motion.
Second, we argued that the approximate efficient spectral representation exists even when the \LIOM have exponentially suppressed tails.

Our results have a number of important implications.
First, in $D=1$, the spectral tensor network can be used to efficiently compute many physical observables, as discussed in more detail in Appendix~\ref{App:onedim} and Appendix~\ref{App:PerfectSampling}.
Second, in systems with s\LIOM, the time evolution operator and the Gibbs ensemble have efficient tensor network representations (see Appendix~\ref{App:UZ}).
Thus, the dynamics starting from matrix product states as well as thermal averages can also be efficiently simulated in such systems.
Third, our formalism naturally applies to the case when the set of q\LIOM is not complete (e.g. with topological order or a single-particle mobility edge). 
The above tensor network then projects onto subspaces with a well-defined value for each of the q\LIOM. 

It has been previously observed that highly excited eigenstates should be representable by tensor networks due to the area law for the scaling of entanglement entropy observed numerically in such states \cite{Bauer:2013rz,Swingle:2013pb,Friesdorf:2014ib}.
Our result is much stronger as it argues for a \emph{single} efficient tensor representing the exponentially many states in the spectrum.
We also emphasize that the assumptions in this article are much weaker than the existence of a quasi-local unitary $U$ that diagonalizes the Hamiltonian of the system.
The efficient spectral tensor network is a trivial consequence if $U$ exists, as the projector onto an energy eigenvector $\ket{E_{\vec{\mu}}}$ is obtained by the action of $U$ on the product state $\ket{\vec{\mu}}$.
The existence of $U$ in MBL systems is however controversial: while there is positive evidence from perturbative work in specific models \cite{Imbrie:2014jk}, numerical studies also find rare eigenstates with large entanglement entropy that would make $U$ non-local \cite{Bauer:2013rz}.
Our construction of the spectral tensor network thwarts the issue by relying only on the existence of an algebraically independent and orthogonal set of q\LIOM.

Tensor network approaches to ground states have significantly advanced our understanding of quantum many-body phenomena in ergodic systems.
They have provided the language for the classification of gapped phases of ordered quantum matter, and allowed for efficient numerical simulations. 
We envisage that the spectral tensor network built in this paper will similarly play a significant role as a natural formalism to efficiently describe MBL phases.
The key open question is the efficient identification of the set of q\LIOM. 
In a previous work \cite{Chandran:2014xr}, a subset of the authors have considered the \LIOM obtained by time-averaging and argued for their locality. 
Such operators could be usefully truncated to obtain approximate integrals of motion; this is an avenue for future research.

\emph{Note added: } A recent preprint \cite{Friesdorf:2014ib} shows that individual eigenstates of MBL systems can be represented as matrix product states under certain assumptions and is complementary to our work.

\paragraph*{Acknowledgements: }
This work was supported by the John Templeton Foundation and by the Simons Foundation (Many Electron Collaboration).
Research at Perimeter Institute is supported by the Government of Canada through Industry Canada and by the Province of Ontario through the Ministry of Research and Innovation.

\appendix

\section{Efficient computation in $D=1$}
\label{App:onedim}
\begin{figure}[htbp]
\begin{center}
\includegraphics[width=8.5cm]{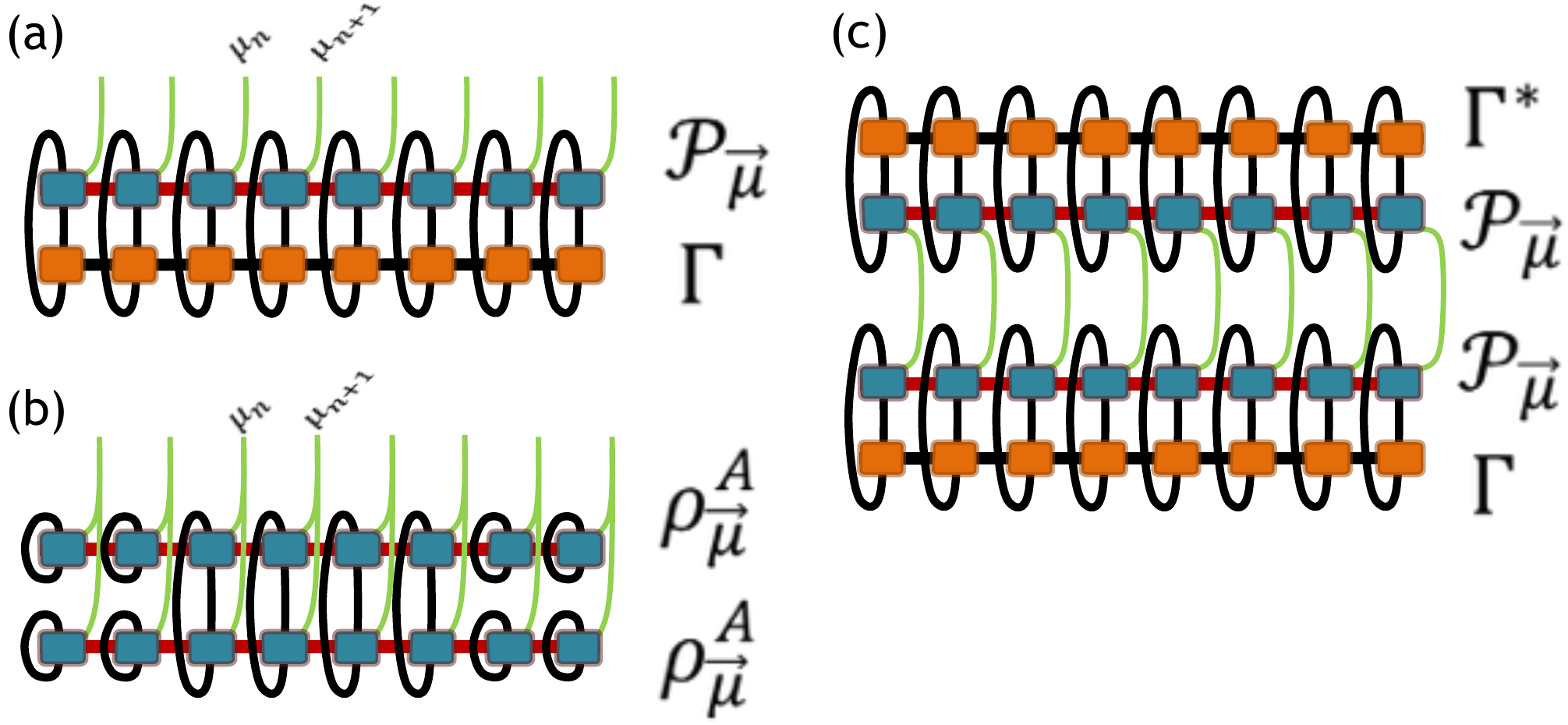}
\caption{ Graphical representations of physical quantities that can be efficiently computed in $D=1$. (a) Expectation value of MPO $\Gamma$ in the energy eigenstate $\ket{E_{\vec{\mu}}}$, (b) Second Renyi entropy of the eigenstate $ R_{2}(\rho^{A}_{\vec{\mu}}) \equiv \tr\left((\rho_{\vec{\mu}}^A)^2 \right)$, (c) Spectral average of the observable $\sum_{\vec{\mu}} |\Gamma_{\vec{\mu}}|^2$. The cost of manipulating these objects is $O(N)$.
}
\label{Fig:observables1d}
\end{center}
\end{figure}

The tensor network of the projector $ \PP_{\vec{\mu}}$ in Fig.~\ref{fig:SpectralTN3} can be used to efficiently compute many physical observables in $D=1$. 
Three examples are shown in Fig.~\ref{Fig:observables1d}.
Fig.~\ref{Fig:observables1d}(a) is the graphical representation of the expectation value of a matrix product operator $\Gamma$ in the energy subspace labelled by $\vec{\mu}$, $\textrm{Tr} (\PP_{\vec{\mu}} \Gamma)$.
Physically relevant $\Gamma$ include local operators like the magnetization, spatially separated operators, non-local string operators, a sum of local operators like the Hamiltonian, a projector onto a matrix product state etc.
We can also efficiently compute observables that are non-linear in $ \PP_{\vec{\mu}}$, such as the second Renyi entropy (Fig.~\ref{Fig:observables1d}(b)).
In addition, we can use the representation in Fig.~\ref{fig:SpectralTN3} to compute an average over the entire energy spectrum by tracing over the $\mu$ indices.
For example, Fig.~\ref{Fig:observables1d}(c) is the graphical representation of the spectral average of the observable $|\bra{\vec{\mu}} \Gamma \ket{\vec{\mu}} |^2$ evaluated in each eigenstate (for simplicity, $N_{\ZZ}=N$). 
Notice that the computation of the average would be inefficient if we had an independent tensor network for each energy eigenstate as there are exponentially many eigenstates; the efficiency here really stems from the spectral tensor network.
In Appendix~\ref{App:PerfectSampling}, we also discuss a different statistical scheme known as perfect sampling to evaluate averages over the entire energy spectrum.

\section{Efficient tensor networks for time evolution and Gibbs ensembles}
\label{App:UZ}
In systems with s\LIOM, the time evolution operator and the Gibbs density matrix also have efficient tensor network representations.
Let the Hamiltonian $H$ be a sum of terms of finite range $H=\sum_i h^{[i]}$.
We define the operator $\tilde{h}^{[i]}$ as the diagonal part of $h^{[i]}$ in the energy eigenbasis:
\begin{equation}\label{eq:hi}
    \tilde{h}^{[i]} \equiv \sum_{\vec{\mu}} \PP_{\vec{\mu}} h^{[i]}  \PP_{\vec{\mu}}.
    \end{equation}
For simplicity, we assume that the number of s\LIOM, $N_{\ZZ}$ is equal to $N$ so that $\PP_{\vec{\mu}}$ is a projector onto a unique energy eigenstate.
Now, $h^{[i]}$ and $\PP^{(n)}_{\mu_n}$ both have finite support.
When their supports have no overlap:
\begin{align}
\sum_{\mu_n} \PP^{(n)}_{\mu_n} h^{[i]} \PP^{(n)}_{\mu_n}  &= h^{[i]} \sum_{\mu_n} \PP^{(n)}_{\mu_n} \PP^{(n)}_{\mu_n} \\
&= h^{[i]}
\end{align}
Thus, Eq.~\eqref{eq:hi} can be simplified to:
\begin{align}
 \tilde{h}^{[i]} = \sum_{\vec{a}} \PP_{\vec{a}} h^{[i]}  \PP_{\vec{a}}.
  \end{align}
where the vector $\vec{a}$ contains the $\mu_n$ of the s\LIOM whose support overlaps with the support of $h^{[i]}$.
The $\tilde{h}^{[i]}$ operators have a number of desirable properties such as: 1) they have finite support on the lattice, 2) they commute with one another on different sites, and 3) the Hamiltonian can be expressed as $H= \sum_{j} \tilde{h}^{[j]}$.
The unitary evolution operator for time $t$, $U(t) \equiv e^{-iHt}$, can thus be decomposed into a product of commuting local evolution operators:
\begin{align}
\label{Eq:UDecomp}
U(t) = \prod_{i} U^{[i]},\quad U^{[i]} \equiv e^{-i \tilde{h}^{[i]}t}, \quad[U^{[i]}, U^{[i']}]=0.
\end{align}
Each $U^{[i]}$ can be efficiently expressed as an MPO as the support of $U^{[i]}$ is finite.
Using Eq.~\eqref{Eq:UDecomp}, we can therefore build an efficient tensor network representation for the time evolution operator up to any time by multiplying the MPO's corresponding to all the $U^{[i]}$'s.

Observe that the argument above applies \emph{mutatis mutandis} to the Gibbs ensemble, $e^{-\beta H}$. 
Thus, the Gibbs ensemble also has an efficient tensor network representation in systems with s\LIOM.

\section{Perfect sampling }
\label{App:PerfectSampling}

\begin{figure}[ht]
\begin{center}
\includegraphics[width=6.5cm]{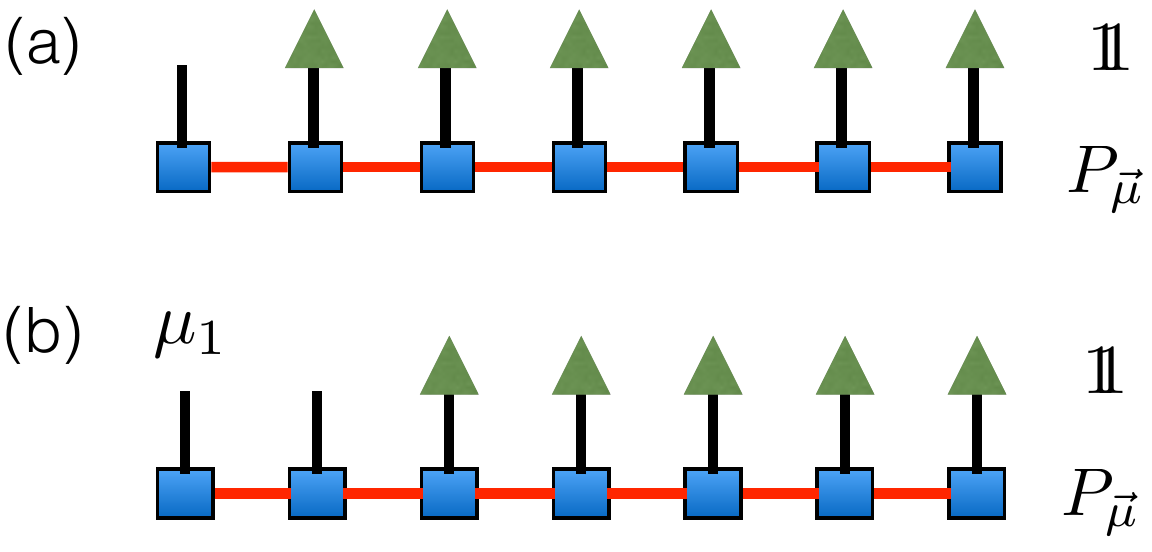}
\caption{ (a) $Q(\mu_1)$, the marginal distribution defined in Eq.~\eqref{Eq:Qmu1}. (b) The conditional probability $Q(\mu_2|\mu_1)$. The green triangles represent unit vectors of length $q$.}
\label{Fig:perfect}
\end{center}
\end{figure}

To evaluate quantities that are non-polynomial functions of the spectrum like connected correlators in $D=1$, perfect sampling is a useful tool.
Let the system be in a mixed state with weight $P({\vec{\mu}})$ for eigenstate $|E_{\vec{\mu}}\rangle$:
\begin{align}
\rho = \sum_{\vec{\mu}} P(\vec{\mu}) \PP_{\vec{\mu}}
\end{align}
Let $P(\vec{\mu})$ have an efficient MPS representation (for example, the Gibbs distribution or the diagonal ensemble of a MPS $\ket{\psi}$). 
Perfect sampling is a method to efficiently draw statistically independent configurations $\vec{\mu}$ according to the probability $P(\vec{\mu})$ at a cost $O(N)$.
Thus, we can generate $M$ independent samples at a cost $O(NM)$ and efficiently approximate quantities like the weighted connected correlator of an observable $O$ in $\rho$:
\begin{align*}
W &= \sum_{\vec{\mu}} P(\vec{\mu}) C_{OO}(\vec{\mu}) \approx \frac{1}{M} \sum_{\vec{\mu} \in \Omega} C_{OO}(\vec{\mu}), 
\end{align*}
where $\Omega$ denotes a set of $M$ uncorrelated configurations $\vec{\mu}$ drawn according to the probability $P(\vec{\mu})$, and $C_{OO}(\vec{\mu})$ is defined as:
\begin{align*}
C_{OO}(\vec{\mu}) &\equiv \bra{E_{\vec{\mu}}} O(x) O(y) \ket {E_{\vec{\mu}}}_c \\
&= \textrm{Tr} (O(x) O(y) \PP_{\vec{\mu}}) - \textrm{Tr} (O(x) \PP_{\vec{\mu}}) \textrm{Tr}( O(y) \PP_{\vec{\mu}} )
\end{align*}
For each $\vec{\mu}$, we can efficiently compute $C_{OO}(\vec{\mu})$, as $O$ and $\PP_{\vec{\mu}}$ have MPO representations.
Perfect sampling generates $\Omega$ (see below), and thus we have an efficient statistical sign problem free scheme to approximate $W$.
The statistical error in the estimate decreases as $1/\sqrt{M}$.

Let us now generate $\Omega$ by adopting the scheme proposed by Ferris \cite{Ferris:2014qa}.
First, define the marginal distribution:
\begin{align}
\label{Eq:Qmu1}
Q(\mu_1,\ldots \mu_i) \equiv \sum_{\mu_{i+1}, \cdots, \mu_N} P(\mu_1,\mu_2,\mu_3,\cdots,\mu_N).
\end{align}
We begin by computing $Q(\mu_1)$ by tracing over $\mu_2 \ldots \mu_N$, as shown in Fig.~\ref{Fig:perfect} (a). We draw a value $\mu_1$ according to this distribution. Fixing the first index to be this value, we then compute the conditional probability of $\mu_2$ given $\mu_1$, $Q(\mu_2|\mu_1)$, by summing over $\mu_3, \ldots \mu_N$ (see Fig.~\ref{Fig:perfect} (b)). We then draw a value $\mu_2$ according to $Q(\mu_2|\mu_1)$. Note that as $Q(\mu_1,\mu_2) = Q(\mu_1)Q(\mu_2|\mu_1)$, we have randomly drawn a pair of value $(\mu_1, \mu_2)$ according to the probability $Q(\mu_1, \mu_2)$. We now repeat: we fix the first two indices to be the values we picked, compute $Q(\mu_3|\mu_1,\mu_2)$ etc. After $N$ such repetitions, we have randomly drawn a vector $\vec{\mu}$ with probability $P(\vec{\mu})$, as intended. By properly recycling partial sums on the MPS, this can be accomplished at a cost $O(N^2q^3\chi)$, where $\chi$ is the bond dimension of $P(\vec{\mu})$ assumed to be greater than $q$.

\section{Bond dimension vs $L$ for a quasi-local operator}
\label{App:mbvsL}
\begin{figure}[htbp]
\begin{center}
\includegraphics[width=6.5cm]{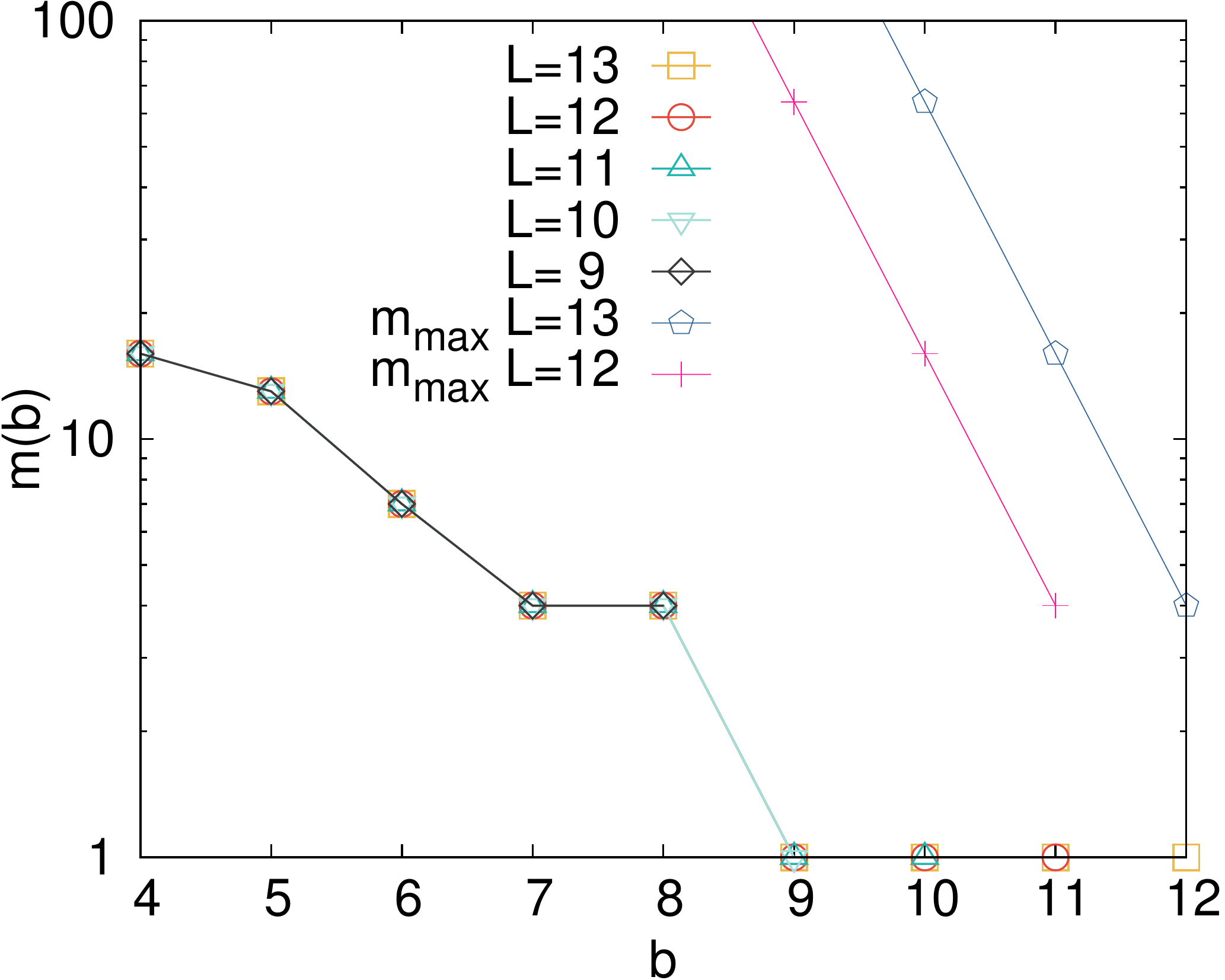}
\caption{Bond dimension of the MPO representation of $e^{iH_{XYZ}} \sigma_{4}^z e^{-iH_{XYZ}}$, $m(b)$ versus the bond index $b$ for truncation parameter $f=10^{-3}$ at different $L$. The maximum value of the bond dimension at $L=12,13$ is also shown for comparison.}
\label{Fig:Ldep}
\end{center}
\end{figure}

In the spin-$1/2$ system described in the numerical section in the main text, we constructed a quasi-local operator by conjugating $\sigma_4^z$ by a quasi-local unitary, $e^{iH_{XYZ}}$. 
In this appendix, we explore the system size dependence of the bond dimension of the MPO representation of $\ZZ= e^{iH_{XYZ}} \sigma_{4}^z e^{-iH_{XYZ}}$.
Shown in Fig.~\ref{Fig:Ldep} is the bond dimension of the MPO representation of $\ZZ$ versus bond index for different $L$ at $f=10^{-3}$.
The pink and blue lines are the maximum possible bond dimensions at $L=12,13$. 
We see that the bond index is small and completely independent of $L$.
This is numerical evidence for the efficient representation of a quasi-local operator by a MPO in the thermodynamic limit up to any required precision.

\bibliography{master}

\end{document}